\documentclass{moriond}

\usepackage[symbol]{footmisc}
\usepackage{amsmath, amssymb}

\def\Journal#1#2#3#4{{#1} {\bf #2}, #3 (#4)}

\def\PRL{\em Phys. Rev. Lett.}
\def\PRD{{\em Phys. Rev.} D}

\newcommand{\beq}{\begin{eqnarray}}
\newcommand{\eeq}{\end{eqnarray}}
\newcommand{\beqnn}{\begin{eqnarray*}}
\newcommand{\eeqnn}{\end{eqnarray*}}

\newcommand{\QCD}{\mathrm{QCD}}
\newcommand{\sphal}{\mathrm{Sphal}}

\def\be{\begin{equation}}
\def\ee{\end{equation}}
\def\bea{\begin{eqnarray}}
\def\eea{\end{eqnarray}}

\newcommand{\Photo}{}

\begin{document}
\vspace*{4cm}
\title{Non-perturbative determination of the $N_f=2+1$ QCD sphaleron rate}

\author{C. Bonanno\textsuperscript{\textit{a}}, F. D'Angelo\textsuperscript{\textit{b}} \footnote[2]{Speaker}, M. D'Elia\textsuperscript{\textit{b}}, M. Naviglio\textsuperscript{\textit{b,c}}, L. Maio\textsuperscript{\textit{d}}}

\address{\textsuperscript{\textit{a}}Instituto de F\'isica Te\'orica UAM-CSIC c/ Nicol\'as Cabrera 13-15 \\ Universidad Aut\'onoma de Madrid, Cantoblanco, E-28049 Madrid, Spain\\\textsuperscript{\textit{b}}Dipartimento di Fisica dell'Università di Pisa \& INFN Sezione di Pisa\\
Largo B. Pontecorvo 3, I-56127 Pisa, Italy\\\textsuperscript{\textit{c}}Scuola Normale Superiore, Piazza dei Cavalieri 7, I-56126 Pisa, Italy
\\\textsuperscript{\textit{d}}Aix Marseille Univ, Universit\'{e} de Toulon, CNRS, CPT, Marseille, France}

\maketitle

\abstracts{The strong sphaleron rate, i.e., the rate of real time QCD topological transitions, is a key phenomenological quantity, playing a fundamental role in several physical contexts. In heavy-ion collisions, a non-vanishing rate can lead to the so-called Chiral Magnetic Effect. In early-Universe cosmology, instead, it can be related to the rate of thermal production of QCD axions. In this talk, we present the first reliable fully non-perturbative computation of the strong sphaleron rate in $N_f=2+1$ QCD at the physical point by means of lattice simulations, in a range of temperatures going from 200 MeV to 600 MeV. Our strategy is based on the inversion of lattice correlators via a recently-proposed modified version of the Backus-Gilbert method.}

\section{Introduction}

The structure of the QCD vacuum is highly non-trivial. The space of gauge configurations has a topological structure meaning that it is divided into homotopy classes and it is not possible to pass from one sector to another by means of continuous transformations without going through infinite energy configurations. Each topological sector is characterized by an integer number, the \textit{topological charge}, defined as the integral in the space-time volume of the \textit{topological charge density} $q(x)$:
\beq
	Q=\int d^4x\ q(x)\ ,\qquad q(x)=\frac{1}{32\pi^2}\epsilon_{\mu\nu\rho\sigma}\mathrm{Tr}\{G^{\mu\nu}(x)\tilde{G}^{\rho\sigma}(x)\}\ .
\eeq

Transitions between different topological sectors occur and they are related to the non-conservation of the axial charge via the ABJ anomaly. More precisely, the topological potential barrier can be overcome through tunneling processes associated to Euclidean instantons. However, when the temperature is sufficiently high, also thermal transitions can happen and these are related to another kind of topological object, the \textit{sphaleron}.

The \textit{strong sphaleron rate} is the rate of real time topological transitions in finite temperature QCD. It is defined as the integral, in the Minkowskian space-time ($t_\mathrm{M}$), of the 2-point correlation function of the topological charge density operator:
\beq\label{eq:sphal_rate_def}
	\Gamma_\sphal = \underset{t_{\mathrm{M}}\to\infty}{\underset{V_s\to\infty}{\lim}} \, \frac{1}{V_s t_{\mathrm{M}}}\left\langle\left[\int_0^{t_{\mathrm{M}}} d t_{\mathrm{M}}' \int_{V_s} d^3x \, q(t_{\mathrm{M}}', \vec{x})\right]^2\right\rangle\ .
\eeq
This quantity has an important phenomenological role both for axion and Quark-Gluon Plasma (QGP) physics. Indeed, on one hand, it is related to the axion thermal production in the early Universe~\cite{axion_sphal}. On the other hand, a non-zero value of $\Gamma_\sphal$ gives rise to local imbalances in the number of left/right-handed quarks in the QGP and consequently to the Chiral Magnetic Effect~\cite{chiral_mag_eff}, i.e., the appearance of an electric current in the direction of a background magnetic field; this point is extremely interesting for experiments involving heavy ion collisions.

Being a topological object, the sphaleron dynamics is purely non-perturbative and it can be investigated, for instance, by means of Monte Carlo simulations on the lattice. However, the lattice framework requires an Euclidean formulation of the theory and Eq.~\ref{eq:sphal_rate_def} can not be used.

In the following, we will talk about the first lattice determination of the strong sphaleron rate in $N_f=2+1$ QCD at the physical point in the temperature range $200~\mathrm{MeV}\lesssim T\lesssim 600~\mathrm{MeV}$. For a more detailed discussion, we refer the reader to the main papers~\cite{main_paper_gauge,main_paper_fullqcd}. 

\section{Numerical setup}

We clarify how $\Gamma_\sphal$ can be extracted from lattice simulations. The strong sphaleron rate is related to the \textit{spectral density} $\rho(\omega)$ of the Euclidean topological charge density time correlator $G(t)$ via the following \textit{Kubo formula}:
\beq\label{eq:sphaleron_from_rho}
	\Gamma_\sphal=2T\lim_{\omega\to0}\frac{\rho(\omega)}{\omega}\ .
\eeq
$\rho(\omega)$ is related to $G(t)$ via an integral relation:
\beq\label{eq:correlator_definition}
	G(t)\equiv\int d^3x \langle q(t,\vec{x})q(0,\vec{0})\rangle = -\int_0^\infty \frac{d\omega}{\pi}\rho(\omega) \frac{\cosh\left[\frac{\omega}{2T}-\omega t\right]}{\sinh\left[\frac{\omega}{2T}\right]}\ ,
\eeq
where now $t$ is the imaginary time.

On the lattice, after choosing a discretization $q_L(x)$ of the topological charge density operator, one can easily compute the correlator $G_L(t)$. However, this observable is affected by short distance singularities that give rise to diverging additive renormalizations in the continuum limit that overcome the physical signal. A \textit{smoothing procedure} is needed to damp all UV fluctuations at energy scales above $1/r_s$, with $r_s$ being the so-called \textit{smoothing radius}.

Once the correlator $G_L(t)$ has been computed, one has to invert Eq.~\ref{eq:correlator_definition}~\footnote[1]{This is an ill-posed problem in the sense that if we have two very close points in the \textit{output space}, the corresponding inputs can be very different so one has to pay attention especially when the outputs are affected by uncertainties.} in order to recover $\rho(\omega)$ and finally the sphaleron rate. To achieve this, we adopt a modified version of the Backus--Gilbert method~\cite{hlt_bgilbert} that allows to express the spectral density as a linear combination of the different time points of the correlator, with the coefficients $g_t$ determined by minimizing a suitable functional. The end of the story is that one can compute the sphaleron rate in the following way:
\beq
	\frac{\Gamma_\sphal}{2T}=\left[\frac{\bar{\rho}(\bar{\omega})}{\bar{\omega}}\right]_{\bar{\omega}=0}=-\pi\sum_{t=0}^{1/T}g_t(0)G_L(t)\ ,
\eeq
and we refer the reader to the main paper for all the details regarding the inversion of the correlator.

\section{Numerical results}

\begin{figure}[!htb]
	\centering
	\includegraphics[scale=0.35]{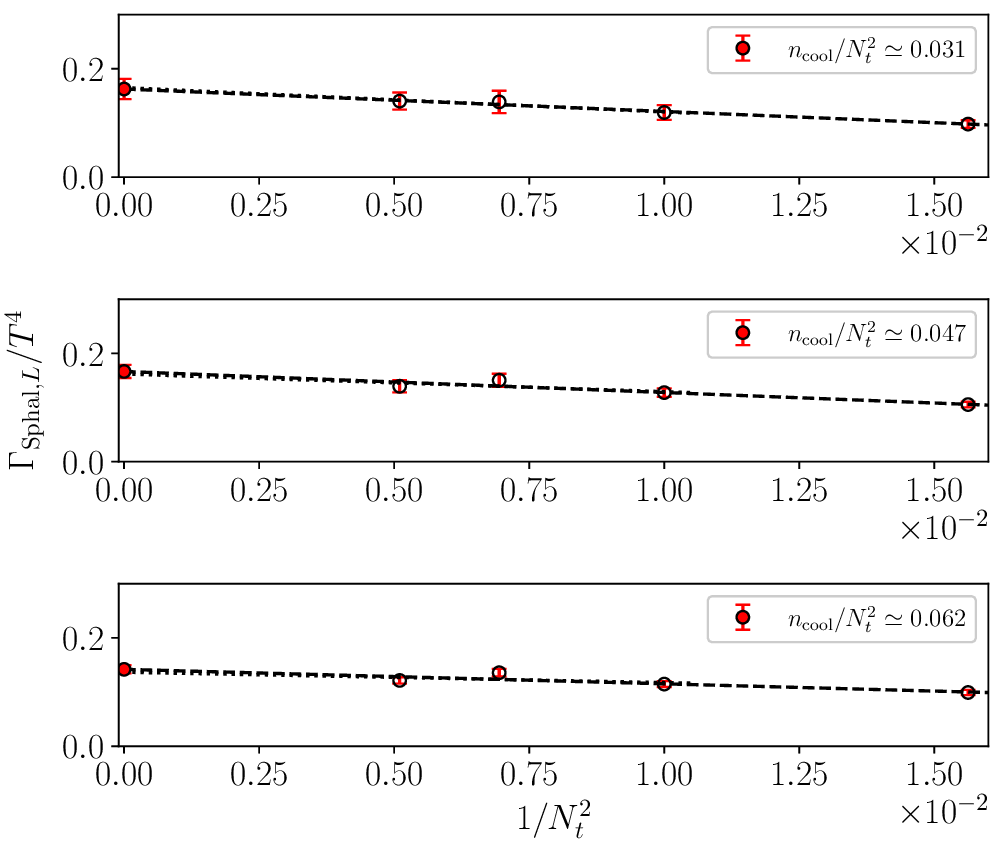}
	\hspace{10mm}
	\includegraphics[scale=0.35]{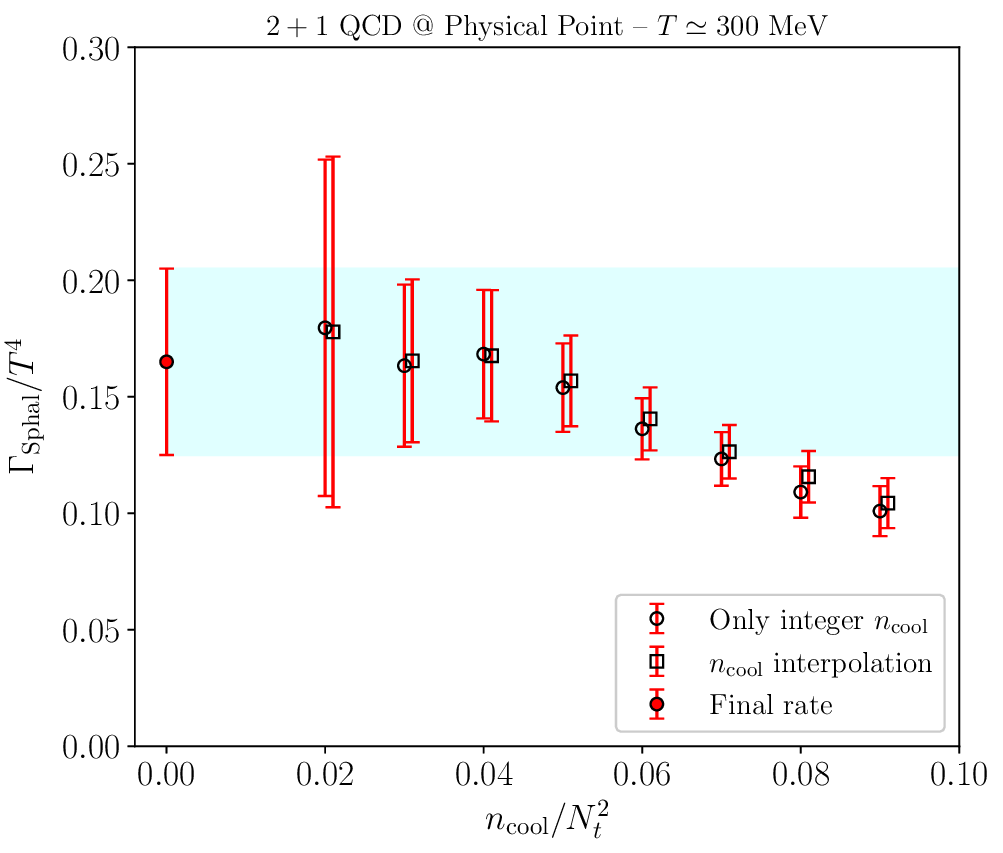}
	\caption{Left: continuum limit of $\Gamma_\sphal/T^4$ for different values of the squared smoothing radius (expressed as $n_{\mathrm{cool}}/N_t^2$, we refer the reader to the main paper for the details). Right: behaviour of the continuum-extrapolated $\Gamma_\sphal/T^4$ as a function of the squared smoothing radius. The shaded band (and the filled circled point) refers to the final result.}
	\label{fig:300mev_result}
\end{figure}

We first invert the correlator at fixed lattice spacing $a$ and smoothing radius $r_s$ $G_L(t;a,r_s)$ to extract the sphaleron rate $\Gamma_\sphal(a, r_s)$. Then we perform the continuum limit at fixed physical smoothing radius and finally the dependence of $\Gamma_\sphal$ on $r_s$ is removed by evaluating the zero-smoothing behaviour of the observable. This procedure is shown in Fig.~\ref{fig:300mev_result} for the temperature $T\simeq300~\mathrm{MeV}$.

As we can notice from the left hand side plot of Fig.~\ref{fig:300mev_result}, the continuum scaling of $\Gamma_\sphal/T^4$ is not affected by large lattice artifacts. Furthermore, the dependence of the continuum value on the smearing radius exhibits a plateau for small $r_s/T\propto\sqrt{n_{\mathrm{cool}}}/N_t$ (Fig.~\ref{fig:300mev_result}, right). This has been taken as the final result for the sphaleron rate: $\Gamma_\sphal/T^4(T=300~\mathrm{MeV})=0.165(40)$.

\subsection{Temperature dependence of $\Gamma_\sphal$ in $N_f=2+1$ QCD}

\begin{figure}[!htb]
	\centering
	\includegraphics[scale=0.35]{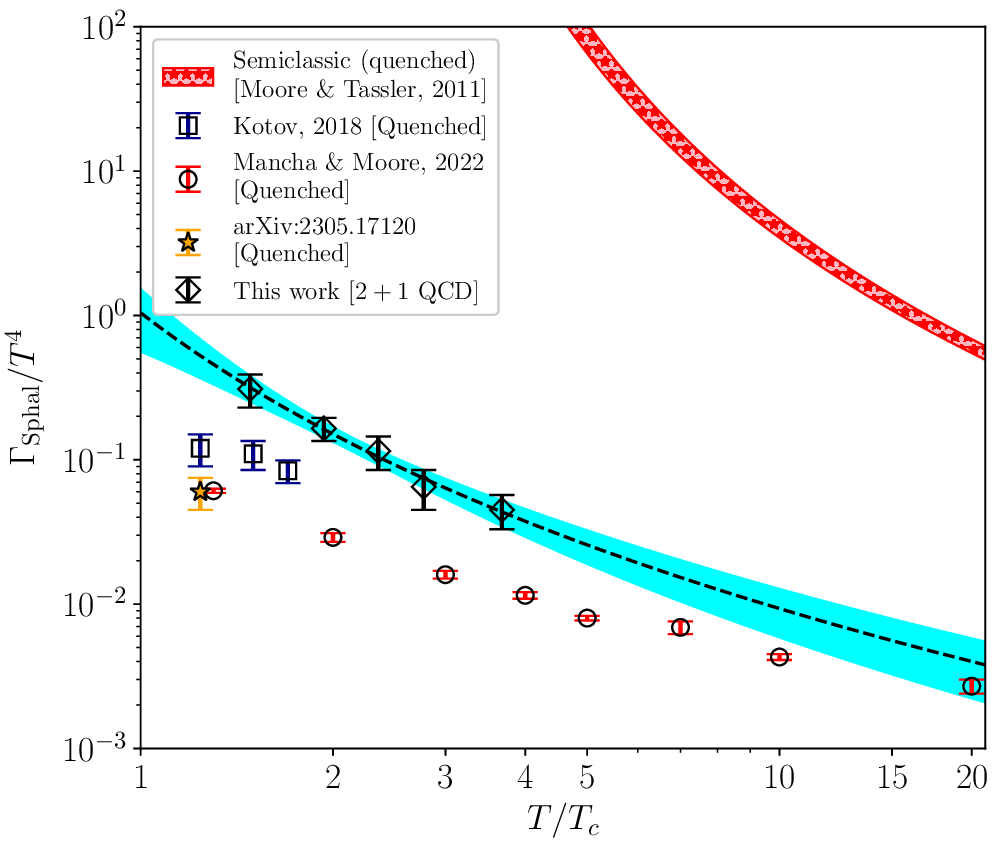}
	\hspace{2mm}
	\includegraphics[scale=0.35]{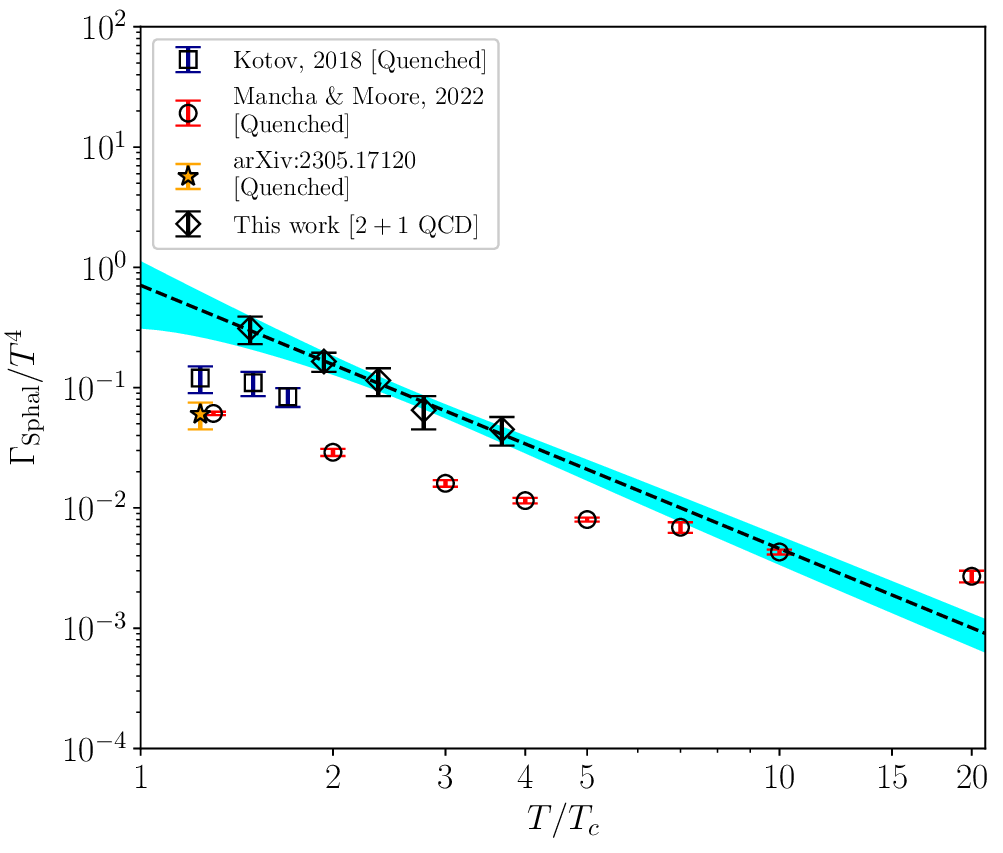}
	\caption{Temperature behaviour of the sphaleron rate. The x-axis is expressed in terms of $T/T_c$, where $T_c=155~\mathrm{MeV}$ for full QCD and $T_c=287~\mathrm{MeV}$ for the quenched theory. The meaning of all symbols is reported in the legend. Left: fit performed with Eq.~\ref{eq:semiclassical_fit}. Right: fit performed with Eq.~\ref{eq:powerlaw_fit}.}
	\label{fig:temperature_dependence}
\end{figure}

We repeated this procedure for 5 different temperatures in the range $200~\mathrm{MeV}\lesssim T\lesssim 600~\mathrm{MeV}$. All results are reported in Fig.~\ref{fig:temperature_dependence}.

In the literature so far only determinations of $\Gamma_\sphal$ in the quenched theory~\cite{oth_res_su3} have been reported, see Fig.~\ref{fig:temperature_dependence}. It is worth noticing that $N_f=2+1$ QCD and quenched results fall in a similar region. This is an extremely interesting point. In the case of instanton-like transitions, it is well known that the fermion dynamics is responsible of the suppression of topological fluctuations: consider the drop in the $T=0$ topological susceptibility $\chi$ from pure SU(3) gauge theory to QCD, being $\chi$ dominated by the short-distance behaviour of $G(t)$. On the other hand, $\Gamma_\sphal$ is dominated by the long-distance tails of $G(t)$, which are not suppressed by light quarks.

We also compared our results with available analytical predictions in literature. If one uses the semiclassical estimation for the sphaleron rate~\cite{semicl} and the 1-loop result for the temperature running of $\alpha_s(T)$~\cite{alpha_s}, one has ($A_0=C_1^{1/5}C_2$, $B_0=T_c/\Lambda_{\mathrm{QCD}}$):
\beq\label{eq:semiclassical_fit}
	\frac{\Gamma_\sphal}{T^4}\simeq C_1\alpha_s(T)^5 = C_1\left[\frac{C_2}{\log(T^2/\Lambda_{\mathrm{QCD}}^2)}\right]^5 = \left[\frac{A_0}{\log(T^2/T_c^2)+\log(B_0^2)}\right]^5\ .
\eeq
We tried to perform a fit according to this function where we left the exponent as a free parameter ($C$). If we fix $C=5$, we obtain $\tilde{\chi}^2/ndof=0.36/3$, $A=2.96(51)$, $B=4.3(1.7)$ that have to be compared with the theoretical expectations $A_0=3.08(2)$ and $B_0=0.46(2)$ (if we use $T_c=155~\mathrm{MeV}$ and $\Lambda_\QCD^{\mathrm{(\overline{MS})}}(\mu=2~\mathrm{GeV})=338(12)~\mathrm{MeV}$~\cite{flag}). If $C$ is left as a free parameter, we obtain a good $\tilde{\chi}^2/ndof$ but a value of $C=5.6$ with a 100\% error.

We also tried to fit our data with a simple power law behaviour:
\beq\label{eq:powerlaw_fit}
	\frac{\Gamma_\sphal}{T^4}\simeq\tilde{A}\left(\frac{T}{T_c}\right)^{-b}
\eeq
\vspace*{-1em}
obtaining $\tilde{\chi}^2/ndof=0.48/3$, $\tilde{A}=0.71(23)$ and $b=2.19(38)$.

\section{Conclusions}

We presented the first computation of sphaleron rate for $N_f=2+1$ QCD at the physical point from lattice simulations in the temperature range going from 200 to 600 MeV. Our data are well described by a logarithmic power-law decay but also by a regular power-law behaviour. It is clear that further investigation is needed and that a larger temperature range has to be studied. 

It would be interesting to extend the present study up to $T\simeq1~\mathrm{GeV}$, and  to the $\vert \vec{k} \vert \neq 0$ case, that is crucial for axion thermal production~\cite{axion_sphal}.

\section*{Acknowledgements}

It is a pleasure to thank G.~Gagliardi, V.~Lubicz, F.~Sanfilippo and G.~Villadoro for useful discussions. The work of C.~Bonanno is supported by the Spanish Research Agency (Agencia Estatal de Investigación) through the grant IFT Centro de Excelencia Severo Ochoa CEX2020-001007-S and, partially, by grant PID2021-127526NB-I00, both funded by MCIN/AEI/10.13039/50110001 1033. C.~Bonanno also acknowledges support from the project H2020-MSCAITN-2018-813942 (EuroPLEx) and the EU Horizon 2020 research and innovation programme, STRONG-2020 project, under grant agreement No 824093. Numerical simulations have been performed on the \texttt{MARCONI} and \texttt{Marconi100} machines at CINECA, based on the agreement between INFN and CINECA, under projects INF22\_npqcd and INF23\_npqcd.

\end{document}